\documentclass[11pt]{article}
\usepackage{graphicx}
\usepackage{authblk}
\usepackage{url}
\usepackage{CJKutf8}
\usepackage{newtxtext,newtxmath}

\title{Development and Validation of the Japanese Moral Foundations Dictionary}
\author[1]{Akiko Matsuo}
\author[2,5]{Kazutoshi Sasahara\thanks{Correspondence should be addressed to \texttt{sasahara@nagoya-u.jp}}}
\author[3]{Yasuhiro Taguchi}
\author[4]{Minoru Karasawa}
\affil[1]{Department of Psychology, Graduate School of Environmental Studies, Nagoya University, Nagoya, Japan}
\affil[2]{Department of Complex Systems Science, Graduate School of Informatics, Nagoya University, Nagoya, Japan}
\affil[3]{Department of Humanities and Social Sciences, School of Letters, Aichi University, Toyohashi, Japan}
\affil[4]{Department of Cognitive and Psychological Sciences, Graduate School of Informatics, Nagoya University, Nagoya, Japan}
\affil[5]{JST, PRESTO, Kawaguchi, Japan}
\date{}

\begin{document}
\maketitle
\begin{abstract}
The Moral Foundations Dictionary (MFD) is a useful tool for applying the conceptual framework developed in Moral Foundations Theory and quantifying the moral meanings implicated in the linguistic information people convey.
However, the applicability of the MFD is limited because it is available only in English.
Translated versions of the MFD are therefore needed to study morality across various cultures, including non-Western cultures. The contribution of this paper is two-fold.
We developed the first Japanese version of the MFD (referred to as the J-MFD) using a semi-automated method---this serves as a reference when translating the MFD into other languages.
We next tested the validity of the J-MFD by analyzing open-ended written texts about the situations that Japanese participants thought followed and violated the five moral foundations.
We found that the J-MFD correctly categorized the Japanese participants' descriptions into the corresponding moral foundations, and that the Moral Foundations Questionnaire (MFQ) scores correlated with the frequency of situations, of total words, and of J-MFD words in the participants' descriptions for the Harm and Fairness foundations.
The J-MFD can be used to study morality unique to the Japanese and also multicultural comparisons in moral behavior.
\end{abstract}

\section*{Introduction}
Currently, one of the most active research areas in social and behavioral sciences pertains to how and on what grounds ordinary people form moral judgments.
A central message from this flourishing body of research is that people quickly decide whether a particular act is morally right or wrong; however, it takes them a relatively long time to provide a ``why'' explanation for their judgment~\cite{haidt2001emotional}.
This intuitionist model of moral judgment has produced voluminous empirical research as well as a comprehensive theoretical framework---this is now formulated as the Moral Foundations Theory (MFT).

The central principle of the MFT is that people inherit a limited number of conceptual templates used for their intuitive classification of observed acts that are potentially relevant to morality. Specifically, it is assumed that there are five major moral foundations including:
(1) ``Care,'' which focuses on not harming others and protecting the vulnerable;
(2) ``Fairness,'' which assumes equivalent exchange without cheating to be good;
(3) ``Ingroup,'' which concerns a collective entity instead of individuals, such as family, nation, team, and military;
(4) ``Authority,'' which postulates respect for authority, resulting in maintaining the hierarchy; and (5) ``Purity,'' which involves a feeling of disgust caused by the impure.

The MFT emphasizes that moral foundations meet not only individuals' adaptive need to fit into their community in the ``correct'' ways but also a collective need for the community to increase its unity and win against other groups.
This is how and why moral foundations are typically shared with a high level of consensus among the community members, according to the MFT.
The consensual nature of moral foundations should manifest most visibly in linguistic communication---this can mobilize the community toward solidarity and sanctity.   
In this group process, moral foundations are assumed to show their political aspects and provide a base for mobilizing members toward their collective goals.
To test this hypothesis concerning ``political'' consensuality, Haidt and colleagues analyzed morality-relevant discourse in daily contexts~\cite{graham2009liberals}.
A tool developed for this purpose was the Moral Foundations Dictionary (MFD), which quantifies virtues and vices associated with each moral foundation expressed in written texts~\cite{graham2009liberals}.

The MFD contains a list of words related to one or several moral foundations such as ``killing'', ``justice'', and ``loyal,'' which correspond respectively to the Care, Fairness, and Ingroup foundations.
The usefulness of the MFD has been demonstrated in empirical studies and combined with the use of the LIWC software (Linguistic Inquiry and Word Count program~\cite{pennebaker2001linguistic}).
For instance, in the research above, Graham et al. analyzed church sermons available online and found that liberal preachers were more likely than their conservative counterparts to use words relevant to Care, Fairness, and Ingroup but less likely to  use words relevant to Authority and Purity.
This is consistent with the general trends documented for these ideological camps on different measures~\cite{graham2009liberals}.

Even though the MFD is useful for linguistic analyses of moral foundations, the dictionary is currently available only in English, and thus it is unknown to what extent we can generalize the findings to the linguistic communities outside of the English-speaking world.
This is a problem because it is plausible that the contents---as well as the roles of different moral foundations---would vary across cultures.
For instance, evidence shows that people living in Western cultures tend to emphasize Care and Fairness in their moral judgments, whereas non-Western people tend to rely more on Ingroup and Purity~\cite{graham2011mapping}.
The use of the MFD translated into different languages might reveal similar differences in communication and discourses.
An accumulating body of evidence concerning cultural differences would also be important in the context of criticism about the potential bias in morality research that leans toward the so-called WEIRD (Western, Educated, Industrialized, Rich, and Democratic) cultural samples~\cite{henrich2010most,Jones1627,clifford2015moral}. 
Furthermore, social media platforms provide an excellent arena for analyzing human behavior in a natural setting, and some recent research has successfully applied natural language processing (NLP) to social media data to quantify people's moral behaviors~\cite{sagi2014measuring,mooijman2017resisting}.
Particularly notable are the findings by Dehghani et al., who identified a key role of Purity in social networking~\cite{dehghani2016purity}, and the work by Kaur and Sasahara~\cite{kaur2016quantifying} showing that although Care was the most dominant, Purity was the most distinct moral foundation in online conversations.
Unfortunately, this evidence is also limited to English texts, because there is no publicly available dictionary that can be applied to texts written in languages other than English~\cite{fulgoni2016, ji2015}.

To overcome this limitation, we describe here how we developed a Japanese version of the MFD (J-MFD) using a semi-automated method.
The J-MFD is publicly available online, and hence our methodology can serve as a useful model for further attempts to develop moral dictionaries in other languages.

\section*{Methods}
\subsection*{Strategy for J-MFD Development}
The translation of a dictionary is beset with at least two difficulties. 
It is difficult to maintain consistency between translation outcomes via multiple translators because the accuracy of rendering from one language to another depends on a translator's linguistic ability. 
Moreover, the translated outcomes are subject to change and require constant updates because the actual use of the translated dictionary in text analysis may lead to a better translation and because language itself culturally evolves. 

To resolve the first issue, we took advantage of computational methods and online linguistic resources and corpora. 
This collection of tools and data allowed us to produce as many translations as possible and ensure accurate translations while reducing human errors. 
To resolve the second issue, we released the J-MFD to the public so that researchers worldwide could freely use it and report issues, if any, on our website.
These comments could be used for future updates.

\subsection*{Development of J-MFD}
We translated the original MFD into our J-MFD via two online linguistic resources and two corpora with the aid of computational methods.
The original MFD contains 324 English moral terms with 11 categories corresponding to ``Virtue'' or ``Vice'' (violates); each is associated with one of the five moral foundations (i.e., Care, Fairness, Ingroup, Authority, and Purity) as well with a more general or abstract category of morality (i.e., Morality General)~\cite{graham2009liberals}. 
Care is henceforth denoted as Harm in accordance with the notation of the MFD.
The moral terms consisted of 156 words (e.g., impair) and 168 word stems (e.g., justifi*, which covers justification, justifier, etc.)
There were some words associated with multiple categories such as ``impair'' (Harm Vice and Purity Vice); other words were associated with only a single category, such as ``justifi*'' (Fairness Virtue alone).

\begin{figure}[t]
\includegraphics[clip,width=1.0\textwidth]{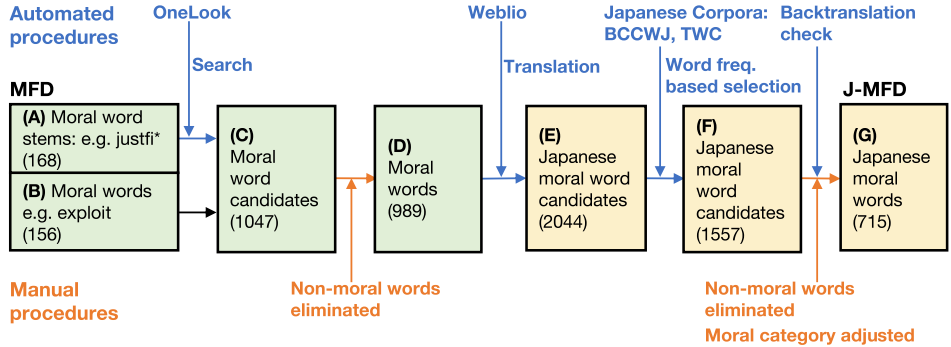}
\caption{Overview of a semi-automated method for J-MFD development. Automated procedures are indicated by downward arrows, and manual procedures are indicated by upward arrows.}
\label{fig:overview}
\end{figure}

Our development followed five steps.
First, our programs automatically collected all words that contained each of the word stems in the MFD by web scraping OneLook---an online dictionary metasearch engine (\url{https://www.onelook.com}) (Fig.~1A$\to$C). 
For example, ``justifi*'' with the ``Filter by commonness: Common words'' option in Onelook returned 11 possible words for the word stem (``justifiable'', ``justifiableness'', ``justifiably'',``justifies'', etc). 
This procedure identified 891 words from all of the word stems in the MFD. 

Next, we manually eliminated 58 words that were unrelated to morality (Fig.~1C$\to$D).
The remaining words comprised a list of moral words for translation.

Third, the remaining words were translated into Japanese via Weblio---an online dictionary and encyclopedia designed for Japanese speakers (\url{https://ejje.weblio.jp}). 
This process was also performed by web scraping, which allowed us to cover possible translation equivalents in Japanese (2044 words) (Fig.~1D$\to$E).

Fourth, we took a frequency-based approach for word selection using two Japanese corpora: Japanese words based on the Balanced Corpus of Contemporary Written Japanese (BCCWJ) and the Tsukuba Web Corpus (TWC). 
BCCWJ is a corpus of contemporary written Japanese that contains 104 million words randomly sampled from books, magazines, newspapers, business reports, blogs, Internet forums, textbooks, and legal documents~\cite{Maekawa2014} (\url{https://pj.ninjal.ac.jp/corpus_center/bccwj/}). 
The TWC is a large corpus that contains 1.1 billion Japanese words obtained from 3.5 million Japanese web pages (\url{http://nlt.tsukuba.lagoinst.info}).
We adopted the top ten most frequent Japanese words for every word stem and the top five for every word in BCCWJ and TWC, thereby filtering out words rarely used (Fig.~1E$\to$F).

Finally, we adjusted the category assignments for each Japanese word after removing words unrelated to morality and words that failed in backtranslation using online dictionaries (Fig.~1F$\to$G).
More specifically, we merged Japanese words whenever possible, using word stem representation. 
For example, \begin{CJK}{UTF8}{ipxm}``違反する''\end{CJK} ({\it Ihan-suru}, or ``violate'') is a verb and \begin{CJK}{UTF8}{ipxm}``違反''\end{CJK} ({\it Ihan}, ``violation'') is its noun form; these words can be merged to \begin{CJK}{UTF8}{ipxm}``違反*''\end{CJK} (n.b. \begin{CJK}{UTF8}{ipxm}``する''\end{CJK} ({\it suru}, ``do'') to make a compound verb).
Similarly, an adjective \begin{CJK}{UTF8}{ipxm}``安全な''\end{CJK} ({\it Anzen-na}, ``safe'') and its noun form \begin{CJK}{UTF8}{ipxm}``安全''\end{CJK} ({\it Anzen}, ``safety'') can be merged to \begin{CJK}{UTF8}{ipxm}``安全*\end{CJK}.''
After the merge procedures, we examined whether Japanese moral word candidates can be back-translated to corresponding English words using online dictionaries.
As a result, 23 words that failed this test were removed, and we were left with 741 Japanese moral terms, for which we adjusted the moral categories.
This adjustment was necessary because multiple words (or word stems) with different moral categories could be translated into the same single Japanese word (or word stem); hence, a single Japanese word could belong to multiple categories. 
Among these categories, the central one (or ones) needed to be selected based on native Japanese knowledge and the definition of moral categories. 
For instance, ``safe*'', ``protect*'', ``shelter'', ``secur*'', ``defen*'', ``guard*'', ``preserve'', and ``obey*'' all can be translated to \begin{CJK}{UTF8}{ipxm}``守る''\end{CJK}; and ``obey*'' can fit in both Authority Virtue and Harm Virtue. 
Because the core meaning of \begin{CJK}{UTF8}{ipxm}``守る''\end{CJK} is a Harm Virtue, we assigned it to Harm Virtue based on the judgment of three native Japanese speakers.

\section*{Results}
\subsection*{Japanese Moral Foundations Dictionary (J-MFD)}
Table~\ref{table:j-mfd_num} shows the number of words for each category and the total number of words in the J-MFD.
As shown, the semi-automated procedures featured more words in Ingroup Virtue and Authority Virtue than in others. 
This seems to reflect Japanese culture, in which group harmony and hierarchy are more appreciated than individual interests.
Example words from the J-MFD are listed in Table~\ref{table:j-mfd_sample}.

\begin{table}[h]
\centering
\caption{The number of words for each category in the J-MFD.}
\label{table:j-mfd_num}
\begin{tabular}{|l|l|l|l|l|l|}
\hline
\begin{tabular}[c]{@{}l@{}}Harm\\ Virtue\end{tabular} & \begin{tabular}[c]{@{}l@{}}Fairness\\ Virtue\end{tabular} & \begin{tabular}[c]{@{}l@{}}Ingroup\\ Virtue\end{tabular} & \begin{tabular}[c]{@{}l@{}}Authority\\ Virtue\end{tabular} & \begin{tabular}[c]{@{}l@{}}Purity\\ Virtue\end{tabular} & \begin{tabular}[c]{@{}l@{}}Morality\\ General\end{tabular} \\ \hline
51 & 42 & 98 & 129 & 89 & 43 \\ \hline
\begin{tabular}[c]{@{}l@{}}Harm\\ Vice\end{tabular} & \begin{tabular}[c]{@{}l@{}}Fairness\\ Vice\end{tabular} & \begin{tabular}[c]{@{}l@{}}Ingroup\\ Vice\end{tabular} & \begin{tabular}[c]{@{}l@{}}Authority\\ Vice\end{tabular} & \begin{tabular}[c]{@{}l@{}}Purity\\ Vice\end{tabular} & Total \\ \hline
93 & 33 & 43 & 53 & 87 & 718 \\ \hline
\end{tabular}
\end{table}

\begin{table}[!h]
\centering
\caption{Examples of moral words in J-MFD}
\label{table:j-mfd_sample}
\begin{tabular}{|p{3cm}|p{4cm}|p{5cm}|}
\hline
 Words in J-MDF ({\it Romaji}) & Counterparts in MFD & Moral Category \\ \hline \hline
\begin{CJK}{UTF8}{ipxm}安全\end{CJK}* ({\it Anzen}*) & safe* & Harm Virtue \\ \hline
\begin{CJK}{UTF8}{ipxm}友情\end{CJK}* ({\it Yujo}*) & amity, fellow*, comrad* & Harm Virtue, Ingroup Virtue \\ \hline
\begin{CJK}{UTF8}{ipxm}殺す\end{CJK} ({\it Korosu}) & kill, kills, killed, killing, destroy & Harm Vice \\ \hline
\begin{CJK}{UTF8}{ipxm}傷\end{CJK} ({\it Kizu}) & hurt*, wound*, stain*, blemish & Harm Vice, Purity Vice \\ \hline
\begin{CJK}{UTF8}{ipxm}平等\end{CJK}* ({\it Byodo}*) & egalitar*, evenness, equal* & Fairness Virtue \\ \hline
\begin{CJK}{UTF8}{ipxm}誠実\end{CJK}* ({\it Seijitsu}*) & constant, honest*, loyal*, integrity & Fairness Virtue, Purity Virtue \\ \hline
\begin{CJK}{UTF8}{ipxm}差別\end{CJK}* ({\it Sabetsu}*) & discriminat*, segregat* & Fairness Vice \\ \hline
\begin{CJK}{UTF8}{ipxm}不法\end{CJK}*({\it Fuho}*) & injust*, lawless*, illegal*, wrong* & Fairness Vice, Authority Vice \\ \hline
\begin{CJK}{UTF8}{ipxm}国民\end{CJK}* ({\it Kokumin}*) & nation* & Ingroup Virtue \\ \hline
\begin{CJK}{UTF8}{ipxm}忠誠\end{CJK}* ({\it Chusei}*) & loyal*, allegian* & Ingroup Virtue, Authority Virtue \\ \hline
\begin{CJK}{UTF8}{ipxm}個人*\end{CJK} ({\it Kojin}*) & individual* & Ingroup Vice \\ \hline
\begin{CJK}{UTF8}{ipxm}不義\end{CJK} ({\it Fugi}) & disloyal*, adulter* & Ingroup Vice, Purity Vice \\ \hline
\begin{CJK}{UTF8}{ipxm}従順\end{CJK}* ({\it Jujun}*) & obedien*, submi*, duti*, complian* & Authority Virtue \\ \hline
\begin{CJK}{UTF8}{ipxm}謙虚\end{CJK}* ({\it Kenkyo}*) & submi*, modesty & Authority Virtue, Purity Virtue \\ \hline
\begin{CJK}{UTF8}{ipxm}扇動\end{CJK}* ({\it Sendo}*) & sedidi*, agitat* & Authority Vice \\ \hline
\begin{CJK}{UTF8}{ipxm}反逆\end{CJK}* ({\it Hangyaku}*) & treason*, traitor*, treacher*, rebel* & Authority Vice, Ingroup Vice \\ \hline
\begin{CJK}{UTF8}{ipxm}きれい\end{CJK}* ({\it Kireina}) & pure*, clean*, pristine & Purity Virtue \\ \hline
\begin{CJK}{UTF8}{ipxm}信心\end{CJK}* ({\it Shinjin}*) & devot*, piety, pious, holy & Purity Virtue, Ingroup Virtue \\ \hline
\begin{CJK}{UTF8}{ipxm}冒涜\end{CJK}* ({\it Botoku}*) & profan*, desecrat* & Purity Vice \\ \hline
\begin{CJK}{UTF8}{ipxm}背教\end{CJK}* ({\it Haikyo}*) & apostate, renegade, pervert & Purity Vice, Ingroup Vice \\ \hline
\begin{CJK}{UTF8}{ipxm}価値\end{CJK}* ({\it Kachi}*) & worth*, value*, good & Moral General \\ \hline
\begin{CJK}{UTF8}{ipxm}正直\end{CJK}* ({\it Shojiki}*) & honest*, integrity, upright, upstanding & Fairness Virtue \\ \hline
\end{tabular}
\end{table}

The J-MFD and a computer program for Japanese word segmentation are publicly available online (\url{{https://github.com/soramame0518/j-mfd/}}). 
Note that Japanese texts are continuous strings of words and are not punctuated by blank spaces; thus, word segmentation is required for Japanese texts before using the J-MFD.

\subsection*{Validation of the J-MFD}
To validate our J-MFD, we compared the mean frequencies of the dictionary words for the five moral foundations that were included in the descriptions about moral issues reported by Japanese participants.
More specifically, 386 Japanese participants (238 men and 148 women; $M_{\rm age} = 35.22$, $SD = 12.30$) were recruited online using the Internet crowdsourcing service Macromill (\url{https://www.macromill.com}). 
Participants read brief explanations of Haidt's five moral foundations (see Supporting Information) and listed as many situations as possible that they thought followed and violated the five moral foundations.

The situations resulted in 16,033 sentences in total after eliminating the responses that were incomprehensible with respect to their meaning or were not related to morality (e.g., ``I can't understand the meaning of the question'').
Each sentence in these descriptions of situations was segmented into words, and five pools of morally relevant words (i.e., Harm-related, Fairness-related, and so forth) were constructed.
For each of these pools, we computed the frequency ratio of appearances of J-MFD words associated with each moral foundation.
To take an example of the word pool produced from the Harm-related context (Virtue and Vice combined), we separately counted the numbers of times that the Harm-related words, Fairness-related words, and so forth contained in the J-MFD appeared in each participant’s descriptions.
To obtain ratio scores, we divided those word counts by the size of the pool and by the total number of dictionary words associated with each moral foundation. 
Figure 2 shows the mean frequency ratio for each foundation in the J-MFD obtained from the Harm-related word pool.
A one-way ANOVA showed a main effect of moral foundations, indicating a significantly higher frequency of Harm words than that of words from the remaining foundations.
We repeated the same analyses for the remaining pools (i.e., Fairness-, Ingroup-, Authority-, and Purity-related), and similarly found the highest frequency ratios in each pool for the corresponding moral foundation, with the main effects of foundation ({\it F} (4,1328) $=$ 52.95, {\it p} $<$ 0.001) (See Figs S1-S4 in Supporting Information).
These results demonstrate the validity of our J-MFD.

\begin{figure}[h]
\includegraphics[clip,width=1.0\textwidth]{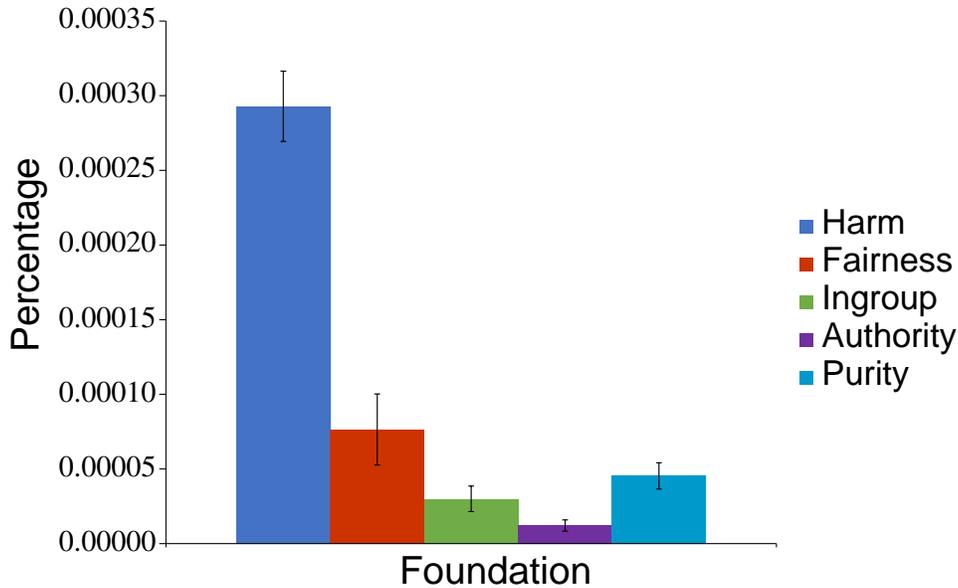}
\caption{Mean percentages of dictionary words per participant used in descriptions about the situations that the participants thought followed and violated the Harm foundation (see Figs. S1-S4 for the other foundations). }
\end{figure}

\subsection*{Relationships between Moral Descriptions and MFQ scores}
We collected self-reported responses to the Moral Foundations Questionnaire (MFQ)~\cite{graham2011mapping} from the same Japanese sample to examine relationships between self-reported moral situations and MFQ scores in Japanese.
This is important to show the applicability as well as the further validation of the J-MFD.
To measure MFQ scores in Japanese, we used the 30-item version of the Japanese MFQ that was back-translated with the approval of the authors of the original MFQ (available at \url{www.moralfoundations.org} and in~\cite{kanai2013}).
The Japanese version of the MFQ was found to have a five factor model as the MFT predicted~\cite{honda2017} and has been used in other research on morality among the Japanese people (e.g.,~\cite{takamatsu2017}).

Our assumption here was that people who have a high MFQ score on a certain moral foundation (e.g., Harm) may have a better-organized schema for the corresponding foundation. 
Thus, when asked to describe situations about the foundation, they could describe it more easily and appropriately than those who have a low MFQ score.
This assumption can be tested by measuring for each of the five moral foundations, (1) how many situations they listed (Virtue and Vice combined), (2) how many words they used to describe the situations, and (3) how many foundation-related words they used to describe the situations.
It should be noted that the task of mentally representing morality gets more specific in the order of (1), (2), and (3): to achieve high performance in (3), a more foundation-specific schema is required for choosing appropriate words related to the corresponding foundation. 
The measurement of (3) with the J-MFD is indispensable for examining the above-mentioned assumption. 
We investigated whether (1), (2), and (3) would correlate with MFQ scores for the corresponding moral foundations.

As for (1), the correlation of MFQ scores and the number of situations described by participants was significant for most of the foundations (Harm: {\it r} $=$ .25, {\it p} $<$ .01; Fairness: {\it r} $=$ .16, {\it p} $=$ .01; Authority: {\it r} $=$ .14, {\it p} $<$ .01; Purity: {\it r} $=$ .22, {\it p} $<$ .01) except the Ingroup foundation ({\it r} $=$ .07, {\it p} $=$ .19).
As for (2), the correlation of MFQ scores and the number of total words included in participant-made situations was significant for all five foundations (Harm: {\it r} $=$ .29, {\it p} $<$ .01; Fairness: {\it r} $=$ .20, {\it p} $<$ .01; Ingroup: {\it r} $=$ .12, {\it p} $=$ .02; Authority: {\it r} $=$ .15, {\it p} $<$ .01; Purity: {\it r} $=$ .22, {\it p} $<$ .01).
As for (3), the correlation of MFQ scores and the number of J-MFD words included in participant-made situations was significant for the Harm and Fairness foundations ({\it r} $=$ .12, {\it p} $=$ .02; and {\it r} $=$ .11, {\it p} $=$ .04, respectively), while there was no significant correlation for the other three foundations (Ingroup: {\it r} $=$ -.02, {\it p} $=$ .67; Authority: {\it r} $=$ .04, {\it p} $=$ .49; Purity: {\it r} $=$ .06, {\it p} $=$ .21).
According to the results of (1)--(3), the correlation between self-reported moral descriptions by Japanese people and MFQ scores was consistent for the Harm and Fairness foundations but not for the other foundations. 
The implications of this finding are discussed in the next section. 

\section*{Discussion}
This work proposed a semi-automated method for translating the Moral Foundations Dictionary (MFD) and developed and validated its Japanese version (J-MFD).
The J-MFD will be updated with revision via collaborative efforts by its users (\url{{https://github.com/soramame0518/j-mfd/}}).
Our method is beneficial for developing other language versions of the MFD, which are needed because multilingual versions allow us to test the Moral Foundations Theory (MFT) in different languages and compare diverse cultures using the same basis of the MFT~\cite{haidt2001emotional,haidt2008morality}.

We showed that the J-MFD allows us to correctly categorize moral-relevant situations in Japanese-written texts into the corresponding moral foundations, which serves as validation of the J-MFD.
Furthermore, our correlation analyses showed that (1) the number of situations, (2) the number of words, and (3) the number of J-MFD words all consistently correlated with the MFQ score in the Harm and Fairness foundations, which implies the existence of a better-developed schema.
People were able to describe Harm- and Fairness-related sentences easily (i.e., (1) and (2)) and accurately (i.e., (3)).
However, such consistent patterns across (1) to (3) were not observed in the Ingroup, Authority, and Purity foundations, which suggests the lack of specific schema for these foundations.
A possible explanation for these results is that Harm and Fairness foundations are more fundamental than the other foundations and are better quantified in the MFQ. 
In contrast, the Ingroup, Authority, and Purity foundations may be more culture-dependent, and the MFQ, as it stands, may inaccurately measure these foundations, and therefore, may need a modification specific to Japanese culture.
These interpretations are consistent with prior research findings, which show that Harm and Fairness are central to moral judgment cross-culturally~\cite{haidt2015, piazza2018} while Ingroup, Authority, and Purity are susceptible to political ideology, ethnicity, culture, and religiosity~\cite{bulbulia2013moral,davis2015,graham2009liberals,johnson2016moral,labouff2017religiosity,yilmaz2016validation}.
Altogether, while our study shows that the J-MFD is a valid tool for morality research in Japanese culture, it also highlights the need for further research to scrutinize the causal relationship of MFQ scores and the use of moral-relevant words in multilingual settings, not only in Japanese.

It is important to note the overrepresentation of male participants in our sample, which featured 238 males and 148 females.
Previous research using internationally diverse samples have found that females were more likely to show higher scores on the Harm, Fairness, and Purity foundations than males, with males scoring higher than females on the Ingroup and Authority foundations in the MFQ~\cite{graham2011mapping}.
In our sample, females rated higher only on Harm ($M_{\rm women} = 24.67$ vs. $M_{\rm men} = 23.76$), and males showed higher mean scores on Fairness ($M_{\rm women} = 22.28$ vs. $M_{\rm men} = 22.46$), Ingroup ($M_{\rm women} = 19.54$ vs. $M_{\rm men} = 20.55$), Authority ($M_{\rm women} = 20.07$ vs. $M_{\rm men} = 20.79$), and Purity ($M_{\rm women} = 21.17$ vs. $M_{\rm men} = 21.94$).
These inconsistent patterns for Fairness and Purity may have been caused by the lack of female participants.
Also, our results from correlation analyses could be affected by the imbalanced number of males in the sample.
Future research with a gender-balanced ratio of participants would be necessary.

This work contributes to interdisciplinary collaborations in morality research across academic fields.
First, the MFT can be tested using NLP with the J-MFD by analyzing languages that people express online~\cite{goolsby1992cognitive}.
In addition to the word-counting approach conducted in this study, NLP allows the J-MFD to be used for word co-occurrence analysis~\cite{dehghani2017tacit} and latent semantic analysis~\cite{sagi2014measuring,kaur2016quantifying}.
Second, the J-MFD can be used in combination with a Japanese emotion dictionary because specific emotions are often associated with moral judgment~\cite{matsumoto2005}.
Third, morality matters in the business world as well; thus, it is important to investigate morality from the perspective of both leaders/employees in organizations and consumers~\cite{goolsby1992cognitive,reed2007moral}.

Finally, the J-MFD allows future research to scrutinize theoretical frameworks about the standards of moral judgment in various fields of study; this would enrich cross-cultural research by facilitating comparison of morality-related texts in English and Japanese.
We are aware that the J-MFD---as well as the original MFD---is not an inclusive dictionary; thus, we had to combine Virtue and Vice categories for words with low occurrence frequency. 
Both dictionaries might be improved by adding more relevant words~\cite{dehghani2016purity,kaur2016quantifying}. 
The expansion of moral words in the J-MFD is critical to improving accuracy in measuring morality from written texts---that is one of our key future goals.

\section*{Author Contributions}
Conceived and designed the experiments: AM, KS, MK. Analyzed the data: AM, KS, YT. Wrote the paper: AM, KS, MK.

\section*{Funding}
This research was supported by JSPS/MEXT KAKENHI Grant Numbers 15H03446 and JP17H06383 in \#4903, JST PRESTO Grant Number JPMJPR16D6, and JST CREST Grant Number JPMJCR17A4. 
The funders had no role in study design, data collection and analysis, decision to publish, or preparation of the manuscript.


\begin{thebibliography}{10}

\bibitem{bulbulia2013moral}
Joseph Bulbulia, Danny Osborne, and Chris~G Sibley.
\newblock Moral foundations predict religious orientations in new zealand.
\newblock {\em {PLoS ONE}}, 8(12):e80224, 2013.

\bibitem{clifford2015moral}
Scott Clifford, Vijeth Iyengar, Roberto Cabeza, and Walter Sinnott-Armstrong.
\newblock Moral foundations vignettes: A standardized stimulus database of
  scenarios based on moral foundations theory.
\newblock {\em {Behavior Research Methods}}, 47(4):1178--1198, 2015.

\bibitem{davis2015}
Don Davis, Kenneth Rice, Daryl Van~Tongeren, Joshua Hook, Cirleen Deblaere,
  Everett Worthington, and Elise Choe.
\newblock The moral foundations hypothesis does not replicate well in black
  samples.
\newblock {\em {Journal of Personality and Social Psychology}}, 110:56--61,
  2015.

\bibitem{dehghani2016purity}
Morteza Dehghani, Kate Johnson, Joe Hoover, Eyal Sagi, Justin Garten,
  Niki~Jitendra Parmar, Stephen Vaisey, Rumen Iliev, and Jesse Graham.
\newblock Purity homophily in social networks.
\newblock {\em {Journal of Experimental Psychology: General}}, 145(3):366,
  2016.

\bibitem{dehghani2017tacit}
Morteza Dehghani, Kate~M Johnson, Justin Garten, Reihane Boghrati, Joe Hoover,
  Vijayan Balasubramanian, Anurag Singh, Yuvarani Shankar, Linda Pulickal,
  Aswin Rajkumar, et~al.
\newblock Tacit: An open-source text analysis, crawling, and interpretation
  tool.
\newblock {\em {Behavior Research Methods}}, 49(2):538--547, 2017.

\bibitem{fulgoni2016}
Dean Fulgoni, Jordan Carpenter, Lyle Ungar, and Daniel Preo{\c t}iuc-Pietro.
\newblock An empirical exploration of moral foundations theory in partisan news
  sources.
\newblock In {\em LREC}, pages 3730--3736, 2016.

\bibitem{goolsby1992cognitive}
Jerry~R Goolsby and Shelby~D Hunt.
\newblock Cognitive moral development and marketing.
\newblock {\em {The Journal of Marketing}}, pages 55--68, 1992.

\bibitem{graham2009liberals}
Jesse Graham, Jonathan Haidt, and Brian~A Nosek.
\newblock Liberals and conservatives rely on different sets of moral
  foundations.
\newblock {\em {Journal of Personality and Social Psychology}}, 96(5):1029,
  2009.

\bibitem{graham2011mapping}
Jesse Graham, Brian~A Nosek, Jonathan Haidt, Ravi Iyer, Spassena Koleva, and
  Peter~H Ditto.
\newblock Mapping the moral domain.
\newblock {\em {Journal of Personality and Social Psychology}}, 101(2):366,
  2011.

\bibitem{haidt2015}
J~Haidt, J~Graham, and P~Ditto.
\newblock The volkswagen of moral psychology.
\newblock Retrieved from http://www. spsp.
  org/news-center/blog/volkswagen-of-morality, 2015.

\bibitem{haidt2001emotional}
Jonathan Haidt.
\newblock The emotional dog and its rational tail: a social intuitionist
  approach to moral judgment.
\newblock {\em {Psychological Review}}, 108(4):814, 2001.

\bibitem{haidt2008morality}
Jonathan Haidt.
\newblock Morality.
\newblock {\em {Perspectives on Psychological Science}}, 3(1):65--72, 2008.

\bibitem{henrich2010most}
Joseph Henrich, Steven~J Heine, and Ara Norenzayan.
\newblock Most people are not weird.
\newblock {\em Nature}, 466(7302):29--29, 2010.

\bibitem{honda2017}
Shiho Honda, Sayaka Ishimaru, Saki Utsunomiya, Tomonari Yamane, Miyu Oda,
  Kazuhisa Sakamoto, Yoshihiro Ohe, Hitomi Kobayashi, Takumi Arima, and Koji
  Kosugi.
\newblock Considering the relation between the haidt's moral foundations theory
  and japanese morality through developing a new scale.
\newblock {\em {Bulletin of the Faculty of Education, Yamaguchi University}},
  66:95--106, 2017.

\bibitem{ji2015}
Qihao Ji and Arthur~A. Raney.
\newblock Morally judging entertainment: A case study of live tweeting during
  downton abbey.
\newblock {\em {Media Psychology}}, 18(2):221--242, 2015.

\bibitem{johnson2016moral}
Kathryn~A Johnson, Joshua~N Hook, Don~E Davis, Daryl~R Van~Tongeren, Steven~J
  Sandage, and Sarah~A Crabtree.
\newblock Moral foundation priorities reflect us christians' individual
  differences in religiosity.
\newblock {\em {Personality and Individual Differences}}, 100:56--61, 2016.

\bibitem{Jones1627}
Dan Jones.
\newblock A weird view of human nature skews psychologists{\textquoteright}
  studies.
\newblock {\em Science}, 328(5986):1627--1627, 2010.

\bibitem{kanai2013}
Ryota Kanai.
\newblock {\em Nou ni kizamareta moraru no kigen: Hito wa naze zen wo
  motomerunoka [The origin of morality engraved in the brain: Why do people
  pursue goodness?]}.
\newblock Iwanami Publisher, 2013.

\bibitem{kaur2016quantifying}
Rishemjit Kaur and Kazutoshi Sasahara.
\newblock Quantifying moral foundations from various topics on twitter
  conversations.
\newblock In {\em Proceedings of the 2016 IEEE Big Data}, pages 2505--2512,
  2016.

\bibitem{labouff2017religiosity}
Jordan~P LaBouff, Matthew Humphreys, and Megan~Johnson Shen.
\newblock Religiosity and group-binding moral concerns.
\newblock {\em {Archive for the Psychology of Religion}}, 39(3):263--282, 2017.

\bibitem{Maekawa2014}
Kikuo Maekawa, Makoto Yamazaki, Toshinobu Ogiso, Takehiko Maruyama, Hideki
  Ogura, Wakako Kashino, Hanae Koiso, Masaya Yamaguchi, Makiro Tanaka, and
  Yasuharu Den.
\newblock Balanced corpus of contemporary written japanese.
\newblock {\em {Language Resource and Evaluation}}, 48(2):345--371, 2014.

\bibitem{matsumoto2005}
Kazuyuki Matsumoto, Ren Fuji, and Shingo Kuroiwa.
\newblock An algorithm for estimating human emotions by using semantic
  information of words.
\newblock In {\em Proceeding of the Association for Natural Language Processing
  $11^{th}$ Annual Conference}, pages 145--148, 2005.

\bibitem{mooijman2017resisting}
Marlon Mooijman, Peter Meindl, Daphna Oyserman, John Monterosso, Morteza
  Dehghani, John~M Doris, and Jesse Graham.
\newblock Resisting temptation for the good of the group: Binding moral values
  and the moralization of self-control.
\newblock {\em {Journal of Personality and Social Psychology}}, 2017.

\bibitem{pennebaker2001linguistic}
James~W Pennebaker, Martha~E Francis, and Roger~J Booth.
\newblock Linguistic inquiry and word count: Liwc 2001.
\newblock {\em {Mahway: Lawrence Erlbaum Associates}}, 71(2001):2001, 2001.

\bibitem{piazza2018}
Jared Piazza, Paulo Sousa, Joshua Rottman, and Stylianos Syropoulos.
\newblock Which appraisals are foundational to moral judgment? harm, injustice,
  and beyond.
\newblock {\em {Social Psychological and Personality Science}}, pages 1--11,
  2018.

\bibitem{reed2007moral}
Americus Reed, Karl Aquino, and Eric Levy.
\newblock Moral identity and judgments of charitable behaviors.
\newblock {\em {Journal of Marketing}}, 71(1):178--193, 2007.

\bibitem{sagi2014measuring}
Eyal Sagi and Morteza Dehghani.
\newblock Measuring moral rhetoric in text.
\newblock {\em {Social Science Computer Review}}, 32(2):132--144, 2014.

\bibitem{takamatsu2017}
Reina Takamatsu and Jiro Takai.
\newblock Validation of the japanese version of moral expansiveness scale.
\newblock {\em {Japanese Journal of Interpersonal and Social Psychology}},
  17:93--102, 2017.

\bibitem{yilmaz2016validation}
Onurcan Yilmaz, Mehmet Harma, Hasan~G Bah{\c{c}}ekapili, and Sevim Cesur.
\newblock Validation of the moral foundations questionnaire in turkey and its
  relation to cultural schemas of individualism and collectivism.
\newblock {\em {Personality and Individual Differences}}, 99:149--154, 2016.

\end{thebibliography}

\end{document}